# AI5GTest: AI-Driven Specification-Aware Automated Testing and Validation of 5G O-RAN Components

Abiodun Ganiyu*
NextG Wireless Lab
North Carolina State University
Raleigh, USA
aganiyu@ncsu.edu

Pranshav Gajjar†
NextG Wireless Lab
North Carolina State University
Raleigh, USA
pgajjar@ncsu.edu

Vijay K Shah
NextG Wireless Lab
North Carolina State University
Raleigh, USA
vijay.shah@ncsu.edu

## Abstract

The advent of Open Radio Access Networks (O-RAN) has transformed the telecommunications industry by promoting interoperability, vendor diversity, and rapid innovation. However, its disaggregated architecture introduces complex testing challenges, particularly in validating multi-vendor components against O-RAN ALLIANCE and 3GPP specifications. Existing frameworks, such as those provided by Open Testing and Integration Centres (OTICs), rely heavily on manual processes, are fragmented and prone to human error, leading to inconsistency and scalability issues. To address these limitations, we present **AI5GTest** – an AI-powered, specification-aware testing framework designed to automate the validation of O-RAN components. AI5GTest leverages a cooperative Large Language Models (LLM) framework consisting of *Gen-LLM*, *Val-LLM*, and *Debug-LLM*. Gen-LLM automatically generates expected procedural flows for test cases based on 3GPP and O-RAN specifications, while Val-LLM cross-references signaling messages against these flows to validate compliance and detect deviations. If anomalies arise, Debug-LLM performs root cause analysis, providing insight to the failure cause. To enhance transparency and trustworthiness, AI5GTest incorporates a human-in-the-loop mechanism, where the Gen-LLM presents top-k relevant official specifications to the tester for approval before proceeding with validation. Evaluated using a range of test cases obtained from O-RAN TIFG and WG5-IOT test specifications, AI5GTest demonstrates a significant reduction in overall test execution time compared to traditional manual methods, while maintaining high validation accuracy.

*Both authors contributed equally to this research.
†Both authors contributed equally to this research.

## 1 Introduction

The emergence of Open Radio Access Networks (O-RAN) marks a transformative era in the telecommunications industry, bringing with it a host of promises and benefits for modern network infrastructures [40] [39]. O-RAN is designed to address the growing demand for flexible, cost-effective, and scalable solutions in the era of 5G and beyond. By disaggregating the hardware and software components, it enables operators to reduce vendor lock-in, improve network efficiency, and accelerate innovation cycles. O-RAN's modularity and openness allow for the seamless integration of cutting-edge technologies such as artificial intelligence, machine learning, and advanced network analytics, thereby enhancing the intelligence and adaptability of communication systems [6].

At the heart of O-RAN's transformative potential are its open interfaces, modular architecture, and disaggregated design. These characteristics promote interoperability, enabling components from different vendors to work together seamlessly [22]. The open architecture has catalyzed the creation of a diverse ecosystem of vendors, ranging from established corporations to emerging startups, to design and deploy network components with unprecedented flexibility [14]. Such advancements align with broader industry goals of enhancing network performance, reducing operational costs, and accelerating the global adoption of 5G technologies, paving the way for a more dynamic and inclusive telecommunications landscape.

However, the rapid growth and diversification of O-RAN components introduce significant challenges, particularly in the domain of testing and validation [4, 23, 21]. In traditional RAN systems, which featured a tightly controlled ecosystem with fewer vendors, testing processes were comparatively simpler. Security, compliance, and performance could be ensured within a streamlined framework, as the limited number of components allowed for exhaustive testing and integration [41, 21]. In contrast, O-RAN's open architecture and large vendor ecosystem create an inherently more complex testing environment [38, 19]. The need to validate interoperability, security, and performance across a vast and diverse range of components places unprecedented demands on testing methods and resources. Testing must not only ensure compliance with O-RAN ALLIANCE and 3GPP specifications but also provide a scalable and automated approach for evaluating O-RAN implementations. This complexity is further compounded by the sheer volume of components, making existing testing methods time-consuming and insufficient to meet the demands of modern networks.

To address these challenges, there is an urgent need for more automated, efficient, and scalable testing frameworks. Recognizing this, the O-RAN ALLIANCE has established Open Testing and Integration Centres (OTICs) as collaborative hubs for the verification, integration, testing, badging, and certification of disaggregated RAN components [3]. OTICs provide structured environments with common test platforms and practices, enabling vendors to verify functional compliance with O-RAN ALLIANCE and 3GPP standards and ensuring interoperability of disaggregated 5G access infrastructure elements prior to network deployment [13, 2].

***Contributions.*** Despite the establishment of OTICs and other testing frameworks, current O-RAN validation approaches remain highly manual, fragmented, and resource-intensive. The O-RAN Test and Integration Focus Group (TIFG) and Working Group 5 have defined numerous test cases; however, validating these cases requires cross-referencing multiple specifications from O-RAN ALLIANCE and 3GPP [1] This fragmented process forces engineers to manually extract expected procedural flows from these documents and compare them against real-time signaling messages captured from O-RAN components. The result is a time-consuming, error-prone process that demands significant expertise and often leads to inconsistent validation outcomes, especially in multi-vendor environments. To address these gaps, we propose AI5GTest, an AI-powered, specification-aware testing framework that automates the validation of O-RAN components to complement existing OTIC efforts by enhancing the scalability, efficiency, and consistency of O-RAN component validation. AI5GTest leverages open-source Large Language Models (LLMs) with 3GPP and O-RAN specifications to (i) *Automatically generate expected procedural flows for test cases, (ii) Validate signaling messages against established standards, reducing manual overhead and enhancing testing consistency and scalability, and (iii) Identify root causes of test failures by analyzing deviations in execution flows and providing insights for debugging.* This automation reduces the likelihood of human errors and enhances the reliability of O-RAN component validation, ensuring that components from various vendors meet the highest standards of performance, interoperability, and security.

Lastly, AI5GTest significantly shortens the O-RAN testing time (< 1 hour/test case), eliminating the need for several hours per test case execution in case of today's manual testing process, as reported in recent studies [37], which has proven to be a huge barrier in the adoption of O-RAN architecture for 5G deployments worldwide. The framework also incorporates key testing components, including a centralized repository of test cases, a packet analyzer (PCAP Analyzer), and a test orchestrator, providing a scalable and automated testing environment.

The main contributions of this paper are as follows.

• We introduce AI5GTest – an AI-powered, specification-aware testing framework designed to automate the validation of O-RAN components. Its goal is to improve the consistency, efficiency, and scalability of multivendor O-RAN component validation, while significantly reducing the overall testing process time.

• We propose Gen-LLM, a novel test case procedural flow generator that automatically generates expected test execution procedural flows based on 3GPP and O-RAN specifications. Gen-LLM integrates a human-in-the-loop mechanism, where testers are presented with the top-k relevant official specifications for each test case. This allows testers to review and approve the AI-generated procedural flows before validation, ensuring transparency and 100% accuracy with standards.

• We propose Val-LLM, a structured validation algorithm that utilizes LLM operations to systematically verify signaling messages, cross-reference them with expected procedural flows, detect deviations, and ensure compliance with 3GPP and O-RAN standards.

• We develop Debug-LLM, an algorithm designed to identify the root causes of failures by analyzing execution flow discrepancies between observed signaling messages and expected procedural flows. Debug-LLM categorizes test cases as 'Partial Pass' or 'Fail', and provides insight for debugging.

• We evaluate AI5GTest using 24 test cases derived from the O-RAN TIFG and WG5-IOT specifications [28, 26]. See Table 2 for the test case details. Gen-LLM is applied to all 24 cases to generate expected procedural flows. Of these, 12 test cases are executed on an over-the-air 5G O-RAN testbed and used to evaluate the Val-LLM and Debug-LLM components [2]. Additionally, we simulate three failure scenarios based on the "Initial UE Attach" test case (one of the 12 test cases), to assess AI5GTest's ability to identify root causes. In total, 15 packet trace files are included in our evaluation set for the Val-LLM and Debug-LLM. Compared to the manual testing process documented in industry reports [37], AI5GTest significantly reduces execution time, thus enhancing the scalability and acceleration of O-RAN testing efforts. The AI5GTest codebase and all 15 packet trace files are made available at [1].

The rest of the paper is organized as follows. Section 2 reviews the related works. Section 3 offers an overview of the O-RAN architecture and background to O-RAN testing, and section 4 delves into the integral components of AI5GTest, its design, and its implementation. We present an experimental evaluation of AI5GTest in section 5 and finally conclude in Section 6.

## 2 Related Works

Most of the existing research related to O-RAN testing has been focused on security-driven testing frameworks, which primarily focus on vulnerability detection through techniques such as fuzzing, static analysis, and dynamic analysis [43, 32]. These methods typically assume the compliance of the system under test, focusing instead on testing the security robustness of O-RAN systems. While these frameworks have significantly advanced security testing, they do not address procedural validation, multi-standard specification alignment, or dynamic procedural flow generation—all pivotal for ensuring end-to-end conformance in multi-vendor O-RAN systems. Industrial players like Viavi, Spirent, and National Instruments (NI) have contributed to broader testing frameworks and solutions.

[1]While our investigation primarily focuses on test cases referencing the O-RAN ALLIANCE and 3GPP specifications, we acknowledge that some test cases may also refer to other standards bodies, such as ETSI and ITU-T. We believe that the AI5GTest framework can be extended to accommodate these test cases with minor modifications.

[2]The remaining 12 test cases are only used for evaluating the automated flow generation (i.e., Gen-LLM) and were not executed on the testbed due to the limitations of the open-source srsRAN cellular stack used for testbed prototyping, such as, absence of O-DU and O-RU disaggregation.

Among the notable contributions, the Consistent and Repeatable Testing of O-DU Across Continents paper [23] examines O-DU testing setups and procedures across two OTICs. It identifies challenges in achieving consistent and repeatable testing outcomes due to differences in deployment technologies, test equipment, and virtualization setups. The authors propose best practices for achieving consistent and reliable O-DU testing, such as standardizing software versions and configurations. However, the study also acknowledges the limitations of current testing practices, which heavily rely on manual processes and vendor-specific implementations, making it challenging to scale for diverse multi-vendor systems.

Building on this effort, the Open6G OTIC Blueprint [13] proposes a programmable testing infrastructure for O-RAN and 3GPP systems, emphasizing modularity, tenant isolation, and support for diverse device-under-test (DUT) configurations. By leveraging VLAN-based logical topologies and automated configuration, this framework supports diverse device-under-test (DUT) configurations with reduced manual intervention. While this infrastructure lays a strong foundation for O-RAN testing, it highlights the need for future integration of AI/ML-driven workflows to enhance adaptability in response to evolving network requirements.

Among existing industrial efforts, Spirent's O-RAN End-to-End Testing Solution [8] stands out as the closest approach to AI5GTest. Spirent's solution offers a comprehensive testing scope, real-time emulation, pre-built test libraries across multiple domains, and a unified user interface for managing diverse testing scenarios. It provides a holistic view of KPIs, metrics, and logs. While Spirent's E2E Testing Solution represents a significant step toward automated O-RAN testing, it is configuration-driven and lacks a dynamic, specification-aware validation process. Specifically, it does not generate expected procedural flows from standards and does not validate signaling messages against the standard procedures.

While existing frameworks like Spirent's O-RAN E2E Testing and security-driven testing frameworks like ASTRA-5G [18] represent valuable progress in automated O-RAN testing, they fall short in incorporating dynamic, specification-aware validation mechanisms. Additionally, prior procedural testing frameworks suffer from manual cross-referencing and inconsistent validation outcomes, particularly in multi-vendor systems. AI5GTest addresses these critical gaps by introducing the first AI-powered framework that: *Automates procedural validation through LLM-generated expected flows; Dynamically cross-references multi-standard specifications; Validates signaling data chronologically and comprehensively, ensuring procedural and conformance accuracy.*

# 3 Preliminaries

This section briefly discusses the O-RAN architecture, particularly its key components (i.e., O-CU, O-DU, and O-RU), and the various categories of test cases defined by the O-RAN ALLIANCE. For a detailed understanding of O-RAN architecture, do refer to [40].

## 3.1 O-RAN Architecture

*3.1.1. O-RAN Centralized Unit.* The O-CU is a logical node that hosts the Radio Resource Control (RRC), Service Data Adaptation Protocol (SDAP), and Packet Data Convergence Protocol (PDCP) layers. It is further split into two logical subunits: the O-CU-Control Plane (O-CU-CP) and the O-CU-User Plane (O-CU-UP).

- **O-CU-CP:** Hosts the RRC and the control plane part of the PDCP protocol. It manages signaling tasks such as connection management, mobility management, and radio resource control.
- **O-CU-UP:** Hosts the user plane part of the PDCP and the SDAP protocols. It is responsible for user data processing, ensuring efficient transmission and reception between the O-CU and the O-RAN Distributed Unit (O-DU).

The O-CU connects to the O-DU via F1 interface, which is responsible for both control plane (F1-C) and user plane (F1-U) communications.

*3.1.2. O-RAN Distributed Unit.* The O-DU acts as the intermediary between the O-CU and the O-RAN Radio Unit (O-RU), handling real-time operations crucial for low-latency and high-performance network operations. It processes tasks at the following layers: *High-PHY* - digital signal processing functions like modulation and coding. *Media Access Control (MAC)* - scheduling, error correction (HARQ), and resource allocation. *Radio Link Control (RLC)* - segmentation, reassembly, and retransmission of data packets. The O-DU connects to the O-CU via the F1 interface and to the O-RU through the Open Fronthaul (O-FH) interface, which enables multi-vendor interoperability. Its role ensures efficient coordination between the central control functions and edge radio operations.

*3.1.3. O-RAN Radio Unit.* The O-RU is located at the network edge, hosts the Low-PHY layer, and is responsible for Radio Frequency (RF) processing - signal transmission and reception, beamforming, synchronization, and fronthaul transport - supporting communication with the O-DU over the O-FH interface. It converts analog radio signals to digital signals (and vice versa) for transmission over the fronthaul to the O-DU.

## 3.2 O-RAN Testing - A Brief Primer

O-RAN testing ensures that disaggregated network components from multiple vendors operate seamlessly while meeting performance and functional standards. To address the complexities of multi-vendor ecosystems, the O-RAN ALLIANCE Working Groups have developed comprehensive testing specifications, including conformance, interoperability, end-to-end (E2E), and security testing [25, 28, 26, 27]. Open Testing and Integration Centres (OTICs) provide standardized environments for these tests, issuing certificates and badges that validate component reliability and compatibility, facilitating broader adoption of O-RAN solutions [3].

*3.2.1. Conformance Testing.* Conformance testing validates that O-RAN components adhere to the O-RAN ALLIANCE and 3GPP standards. Defined by O-RAN WG4, it focuses on the management plane (M-Plane) using NETCONF and YANG models for reliable configuration and management, the synchronization plane (S-Plane) for timing and frequency alignment, and the user/control planes (U-Plane/C-Plane) for packet processing and signaling between O-RUs and O-DUs[25].

*3.2.2. End-to-End Testing.* End-to-End (E2E) testing [28], as defined by the Testing Integration and Focus Group (TIFG), assesses integrated system performance from user equipment (UE) to the core network. It verifies functional processes such as connection setup, handovers, and resource management while measuring key performance indicators (KPIs) like throughput and latency. E2E testing also examines service delivery (data, voice, video) under various traffic and radio conditions, with additional focus on RIC-enabled optimizations for mobility control and energy efficiency.

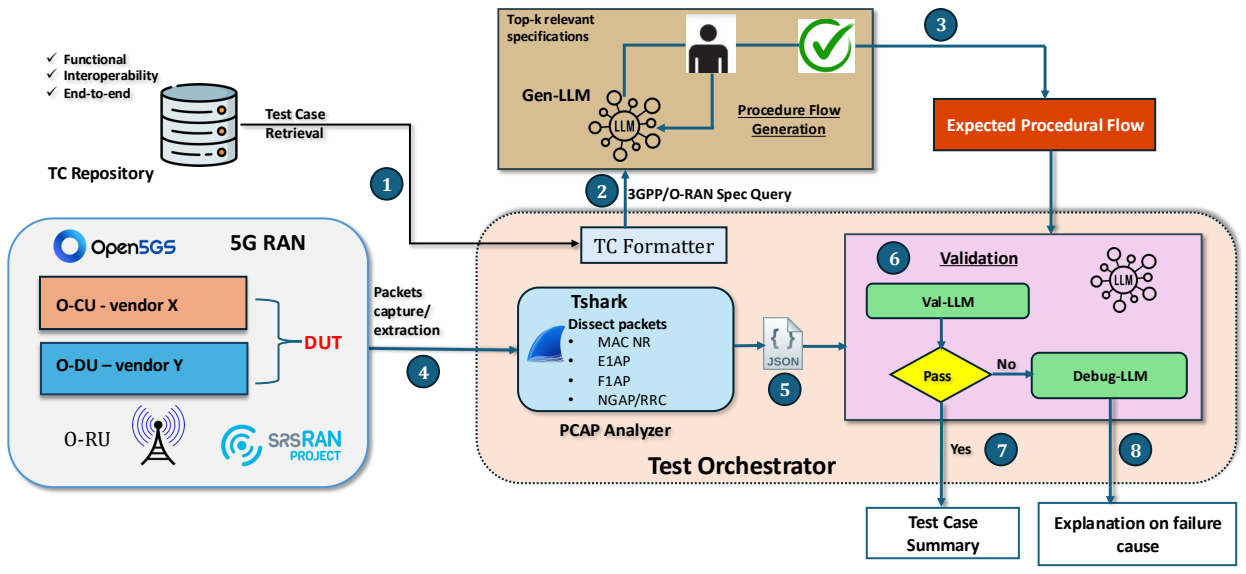

**Figure 1: High-level overview of AI5GTest Framework.**

*3.2.3. Interoperability Testing.* Interoperability testing, defined by O-RAN WG5 [26], ensures multi-vendor components work seamlessly across open interfaces (e.g., W1, E1, X2, F1, and Xn). It validates key interactions such as CU-DU communication on the F1 interface and RU-DU connectivity on the Open Fronthaul interface. E.g., X2 interface testing ensures eNB and en-gNB devices maintain control and user plane connections, supporting reliable handovers and data transmission across different vendor components.

*3.2.4. Security Testing.* Security testing, guided by O-RAN WG11 [27], ensures network resilience against vulnerabilities while maintaining data confidentiality, integrity, and availability. The process validates secure communication protocols (e.g., TLS, IPsec, SSH) and assesses resilience through fuzzing tests and denial-of-service (DoS) simulations. Additionally, mechanisms like Software Bill of Materials (SBOM) and cryptographic signing are employed to ensure the integrity of software components within the O-RAN architecture.

## 4 AI5GTest

This section presents the design of AI5GTest, focusing on its integral components— Test Case Repository, PCAP Analyzer, Test Orchestrator, Gen-LLM, Val-LLM, and Debug-LLM —and how these components interact to provide a unified testing solution to address the key challenges in O-RAN testing and validation.

### 4.1 AI5GTest - A Walkthrough

The entire testing process, from test case initialization to reporting, is visualized in Fig. 1, showcasing the interactions between Gen-LLM, Val-LLM, Debug-LLM, and supporting components.

(1) The Test Orchestrator initiates the process by selecting a predefined test case (e.g., UE attach, inter O-DU mobility) from the Test Case (TC) Repository. Each test case includes essential metadata and descriptions necessary for execution.

(2) The Test Case (TC) Formatter formats the selected test case and generates a structured query describing the expected procedural behavior. This query serves as the input to Gen-LLM for specification-aware flow generation.

(3) Gen-LLM dynamically generates the expected procedural flow by referencing relevant O-RAN and 3GPP specifications. The generated flow outlines the signaling exchanges and protocol behaviors expected for the test case. To ensure accuracy and transparency, the framework incorporates a human-in-the-loop mechanism, where the tester reviews and approves the generated flow. Once approved, the procedural flow is passed to Val-LLM for validation against the observed signaling logs.

(4) During test execution, the PCAP Analyzer analyzes the captured real-time signaling messages exchanged between O-RAN components. The captured Protocol Data Units (PDUs) across multiple 5G layers are dissected and converted into a machine-readable format for validation.

(5) The dissected packet information is structured into a JSON format (as logs representing the observed test procedural flow during testing) and forwarded to the Val-LLM.

(6) Val-LLM compares the observed procedural flow against the expected flow procedure obtained from the Gen-LLM, verifying sequential signaling order and message completeness. If discrepancies are found, Val-LLM flags them for further analysis.

(7) If the DUT successfully passes the test case, the Test Orchestrator compiles a comprehensive test summary. This report details the pass status and confirms procedural compliance.

(8) If the DUT fails the test case, the Test Orchestrator invokes Debug-LLM to perform root cause analysis (RCA) for the observed anomalies. Debug-LLM interprets the nature of failures, references relevant O-RAN and 3GPP specifications, and provides insights for debugging. The debugging process highlights specific deviations, missing or incorrect signaling messages.

### 4.2 AI5GTest Component Details

*4.2.1. Test Case Repository:* It serves as a central repository for storing predefined O-RAN test cases across the various categories of O-RAN test specifications, including interoperability, conformance, and end-to-end testing. Each test case in the repository is accompanied by its detailed description, expected results, and references to the associated O-RAN or 3GPP specification documents. When a test is initiated, the repository retrieves the relevant test case along with all its associated metadata, ensuring that the test execution process is aligned with the required specifications.

*4.2.2. Test Orchestrator:* It acts as the central controller for the AI5GTest framework, ensuring seamless coordination between Gen-LLM, Val-LLM, PCAP Analyzer, and the Debug-LLM. It handles test case selection, triggers packet capture, manages LLM interactions, and compiles final reports. The orchestrator also ensures that each test case execution follows a consistent and reproducible workflow to ensure test validity across diverse multi-vendor O-RAN environments.

*4.2.3. PCAP Analyzer:* It captures and processes real-time signaling messages exchanged between O-RAN components during test execution. Implemented as a Python-based micro-application, it leverages tools like `tcpdump`[3] and `tshark`[4] to capture control plane signaling messages across O-RAN and 5G interfaces, including MAC-NR, F1AP, E1AP, and NGAP. A key enhancement involves automated configuration of Wireshark dissectors for each protocol, addressing the non-trivial decoding of PDUs due to different Data Link Type (DLT) preferences. The analyzed packet data is then structured into standardized JSON files[5] containing essential metadata, ensuring that Val-LLM can efficiently cross-reference observed behaviors with the expected procedural flows generated by Gen-LLM.

[3] A command-line packet analyzer that captures network traffic in real-time, primarily used for network debugging and traffic monitoring [15].
[4] A terminal-based counterpart of Wireshark, enabling advanced packet analysis and protocol dissection [10].
[5] For the remainder of the paper, we refer to these JSON files as log files.

*4.2.4. TC Formatter.* The TC Formatter bridges raw test case descriptions and the procedural flow generation by Gen-LLM. Once a test case is selected from the repository, the formatter processes the associated metadata, including test case descriptions, related interfaces, and referenced specification documents (e.g., 3GPP TS, O-RAN WG TS). It then generates a *structured query* that encapsulates the essential contextual details required by Gen-LLM.

*4.2.5. Gen-LLM.* Gen-LLM is the LLM-driven module responsible for generating expected procedural flows for selected test cases. Built on extensive O-RAN and 3GPP specifications, Gen-LLM processes queries formatted by the TC Formatter and retrieves the top-k relevant specification references for each test case. An important feature of Gen-LLM is its interaction with the tester—before finalizing procedural flows, Gen-LLM presents the top-k specifications for the tester's approval. This step ensures transparency and allows the tester to cross-reference official standards in case of ambiguities. Upon approval, Gen-LLM generates detailed, step-by-step procedural flows in a machine-readable format, ready for validation by Val-LLM. To develop Gen-LLM, we extend ORANSight [12], the state-of-the-art LLM framework for O-RAN applications, by enhancing its retrieval-augmented generation (RAG) capabilities to automate procedural workflow generation.

A key challenge lies in efficiently storing and retrieving contextually relevant information from the vast, heterogeneous documentation of 252 O-RAN (up to Release 4) and 14, 560 3GPP specifications (up to Release 19) while not retrieving tangential or incorrect information to obtain correct procedural flows. To address this, we leverage three main components: a FAISS database [30], an Embedding Generator [42], and a Reranking model [7].

As the atomic task for Gen-LLM can be perceived as identifying the correct specification document that can address a query from the Test Orchestrator, our retrieval process must identify the most relevant and contextually accurate 3GPP/O-RAN specification. Since processing entire documents is computationally infeasible, we address this by first segmenting a document into semantically meaningful chunks, facilitating precise retrieval [33]. These segments ensure that queries access only the most relevant portions of the dataset, reducing unnecessary information and improving retrieval efficiency. To enable effective search, each chunk is transformed into a dense vector representation using a pre-trained embedding model [29]. This model, constrained by a predefined context length, encodes semantic properties of the text, enabling similarity-based retrieval based on contextual relevance. The resulting embeddings are indexed in a specialized database optimized for high-speed searches, allowing for efficient access to pertinent information.

We subject all available O-RAN and 3GPP specifications to this pipeline, forming a comprehensive database that supports procedural flow generation. We employ the `BGE-Large-en-v1.5` [42], a high-dimensional embedding model with a vector size of 1024, to transform text chunks into dense vector representations. These embeddings are indexed in an FAISS database [30], optimized for high-speed similarity searches, ensuring efficient retrieval of relevant document segments. The entire process leads to a database spanning 5, 411, 013 chunks and a total of 573, 325, 641 words.

The LLM we use for RAG is the `Mistral-7B-instruct-v0.3`, which supports a context length of 32k tokens and has been widely leveraged in various open-source LLM applications [17] [11]. These components were specifically chosen due to their demonstrated success in the ORANSight work and to maintain the open-source nature of our solution, aligning with our commitment to transparency and reproducibility. To maintain traceability, metadata such as document references and section numbers are preserved, ensuring a transparent link between retrieved content and official standards. The original RAG pipeline in ORANSight naively ranked documents solely based on the Euclidean distance computed from the embeddings of an input query. This approach did not account for contextual relevance, often retrieving semantically similar yet extraneous content that negatively affected procedural flow generation. Experiments and literature have shown that such a method can lead to low LLM recall [16]. To address this shortcoming, we propose a reranking mechanism using the `BGE-M3` model [7]. This model, capable of processing inputs from short sentences to documents of up to 8,192 tokens, re-evaluates the top-k retrieved chunks to ensure that only the most contextually relevant segments are prioritized, thereby significantly improving the overall accuracy and relevance of the generated procedural flow.

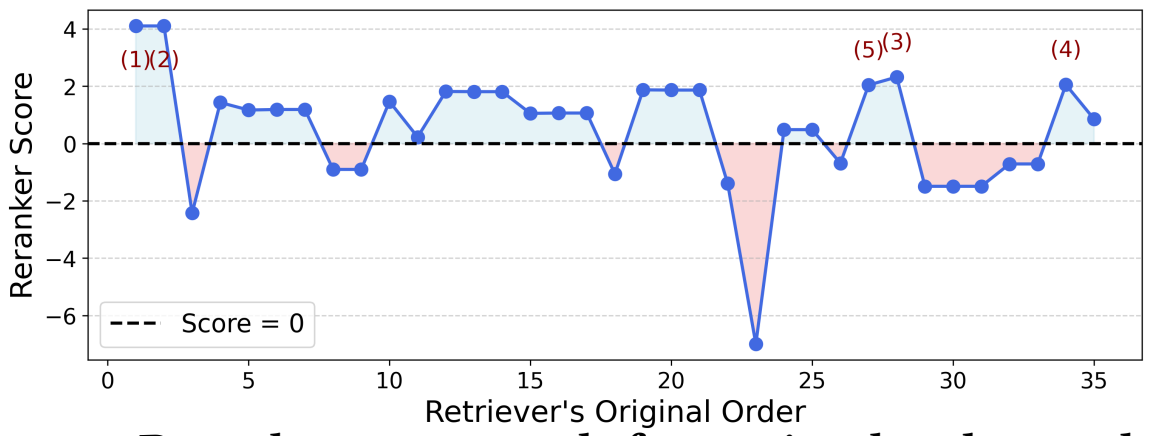

**Figure 2: Reranker score trends for retrieval and procedural flow generation. Blue regions indicate procedurally relevant chunks, while red regions denote extraneous content. Labels (1), (2), (3), ... show the reranker-assigned reordering.**

To demonstrate the difference between our reranker and a naive RAG pipeline, we can consider a taste prompt: *Give the UE Initial Access procedure between gNB-DU, gNB-CU, and AMF.* The Figure 2 shows how the naive pipeline's top-35 retrievals (ordered by embedding distance) frequently receive low or even negative relevance scores when re-evaluated by the `BGE-M3` reranker. Many of these chunks, although semantically similar to the query, are procedurally irrelevant and thus introduce noise into the final output. By contrast, the reranker identifies and prioritizes the segments that are genuinely aligned with the test orchestrator's prompt, effectively reshuffling the naive RAG's ranking. Sample chunks based on the Figure 2 and their reranker scores are illustrated in the Appendix A.3. This selective emphasis on contextually important documents prevents the inclusion of tangential or extraneous information, leading to improved LLM recall and more accurate test case procedural flow generation. For the experiments mentioned in Section 5.4, the Gen-LLM pipeline retrieves a total of 100 documents through the Embedding Generator, out of which the top 15 unique documents obtained through the reranker are used as context to address the TC-formatted query[6]. As the metadata is also saved through the FAISS database, the Gen-LLM can provide the associated specification document for the human-in-the-loop system.

[6]These parameters showcased the best response and further increasing the amount of retrieval resulted in a low LLM recall and performance degradation

*4.2.6. Val-LLM.* The Val-LLM compares the observed behavior during test execution against the expected test case procedure generated by the Gen-LLM. As signaling data and metadata are captured and structured into a machine-readable format by the PCAP Analyzer, the Val-LLM analyzes this data to check for discrepancies, such as deviations from expected message sequences, mismatches in signaling messages, or non-conformance with protocol standards. To realize an accurate system that suffices the Val-LLM roles, experimented with state-of-the-art models such as GPT-4o and Gemini-1.5. However, despite their advanced reasoning capabilities and documented performance in various domains, they are unable to process an entire log file alongside a procedural flow and conclusively determine whether a test case has passed. Both of these models struggle with long-sequence reasoning and with maintaining chronological integrity over extended contexts, and this phenomenon is referred to as *attention overflow* [31].

To address this limitation, we propose a streamlined process that leverages the `LLaMA-3.1: 70B` LLM that is abbreviated as *LLaMA*. We decompose the validation problem into a series of sequential evaluations that guarantee an accurate validation for a given log file and procedural flow. The proposed method processes the log file one entry at a time, maintaining strict chronological order while seeking to identify each step in the procedural flow. The atomic operation of our algorithm is a *LLaMA* forward pass that aims to classify if a particular test case step has been executed in the current log entry, ensuring that *LLaMA* can accurately classify without succumbing to attention overflow. We not only prompt *LLaMA* to classify a particular log entry, but we also instruct it to provide a detailed explanation and a confidence score for the prediction for enhanced interpretability.

The Val-LLM algorithm is depicted in Algorithm 1, and it validates test cases by ensuring the observed behavior in the log file strictly follows the expected procedural flow while operating in a deterministic manner to classify outcomes as either **Pass** or **Fail** without allowing intermediate states. It begins by initializing a step counter $s$ to track the current procedural step and a log index $i$ to iterate through individual log indices $l_i$ in the log file $\mathscr{L}$. For each log entry, we perform a *LLaMA* forward pass to determine if the current procedural step $p_s$ is executed in that index. If the step is identified, we advance to the next step in the procedural flow and move to the next log index, as a single log index cannot satisfy multiple procedural steps. If the step is not found in the current log index, we simply move to the next log index and continue searching. This process continues until either all steps in the procedural flow are found (resulting in a **Pass**) or we reach the end of the log file without finding all steps (resulting in a **Fail**).

*4.2.7. Debug-LLM.* As the Val-LLM can only determine a complete failure and does not support a root cause analysis, such as a missing step or overlooking out-of-order executions of subsequent steps, we propose Debug-LLM, which supports an in-depth analysis at the cost of more LLM runs. Building upon Val-LLM, we extend the approach by performing an exhaustive traversal of the log file $\mathscr{L}$ to identify all executed steps, even when they deviate from the expected procedural flow $\mathscr{P}$. The detailed algorithm is mentioned in

**Algorithm 1** Val-LLM Algorithm

**Require:** Log file $\mathscr{L} = \{l_1, l_2, \dots, l_N\}$, procedural flow $\mathscr{P} = \{p_1, p_2, \dots, p_M\}$
**Ensure:** Classification of test case as {Pass, Fail}
Initialize step counter: $s \leftarrow 1$
Initialize log index: $i \leftarrow 1$
**while** $s \leq M$ **and** $i \leq N$ **do**
    Retrieve current log entry $l_i$
    Perform LLaMA forward pass on $l_i$ to classify if step $p_s$ is executed
    **if** LLaMA classifies $p_s$ as executed in $l_i$ **then**
        **Pass Condition**: Move to next step $s \leftarrow s + 1$
        Increment log index: $i \leftarrow i + 1$
    **else**
        Increment log index: $i \leftarrow i + 1$
    **end if**
**end while**
**if** $s \leq M$ (Not all steps found) **then**
    **Fail Condition**: Mark test case as Fail
**else**
    **Pass Condition**: Mark test case as Pass
**end if**
**Classification:** Return Pass or Fail

2. Unlike Val-LLM, which ensures correct chronology by avoiding already visited log windows for a given step, Debug-LLM performs a forward pass of *LLaMA* for each step $p_s$ across all available log indices $l_i$. The results are stored in $\mathscr{R}$ as tuples containing the step index, the corresponding window, and its classification label.

Once all indices $l_i$ have been analyzed for all the steps, the algorithm extracts the subset $\mathscr{S}$ of executed steps and verifies their chronological integrity by comparing the log window indices associated with consecutive steps; if a later step appears in an earlier window, this indicates an out-of-order execution. Based on this constraint, if all steps in $\mathscr{P}$ are present and in the correct sequence, the test case is classified as a *Pass*, whereas the presence of all steps with any order discrepancy leads to a *Partial Pass*, and the complete absence of a particular step $p_s$ results in a *Fail*.

**Algorithm 2** Debug-LLM Algorithm

**Require:** Log file $\mathscr{L} = \{l_1, l_2, \dots, l_N\}$, procedural flow $\mathscr{P} = \{p_1, p_2, \dots, p_M\}$
**Ensure:** Classification of test case as {Pass, Partial Pass, Fail}
Initialize results list: $\mathscr{R} \leftarrow []$
**for** $s = 1$ to $M$ **do**
    **for** $i = 1$ to $N$ **do**
        Retrieve log entry $l_i$
        Perform LLaMA forward pass on $l_i$ to classify step $p_s$
        Store results: $\mathscr{R} \leftarrow \mathscr{R} \cup \{(s, i, \text{label})\}$
    **end for**
**end for**
**Classification:**
Initialize flag: in_order $\leftarrow$ True
Extract executed steps: $\mathscr{S} \leftarrow \{(s, i) \mid (s, i, \text{label}) \in \mathscr{R}, \text{label} = \text{Executed}\}$
Group $\mathscr{S}$ by step and select earliest log index for each step: $\mathscr{S}' \leftarrow \{(s, \min\{i \mid (s, i) \in \mathscr{S}\}) \mid s \in \{s \mid (s, i) \in \mathscr{S}\}\}$
Sort $\mathscr{S}'$ by step: $\mathscr{S}'' \leftarrow \text{sort}(\mathscr{S}', \text{by} = s)$
**for** $j = 1$ to $|\mathscr{S}''| - 1$ **do**
    Extract $(s_j, i_j)$ and $(s_{j+1}, i_{j+1})$ from $\mathscr{S}''$
    **if** $i_{j+1} < i_j$ **then** ▷ Step $s_{j+1}$ appears earlier than $s_j$
        in_order $\leftarrow$ False
        **Break**
    **end if**
**end for**
**if** in_order = True and $\mathscr{S}''$ contains all steps in $\mathscr{P}$ **then**
    Return Pass
**else if** $\mathscr{S}''$ contains all steps in $\mathscr{P}$ but in_order = False **then**
    Return Partial Pass
**else**
    Return Fail
**end if**

We also visualize the results by plotting the $\mathscr{R}$ in the Appendix. Though the Debug-LLM requires substantially larger runs, we can accurately detect out-of-order steps for a given log file and infer potential root causes of deviations. This visualization provides a comprehensive overview of the test execution, enabling users to pinpoint anomalies and validate the system's compliance with the expected procedural flow.

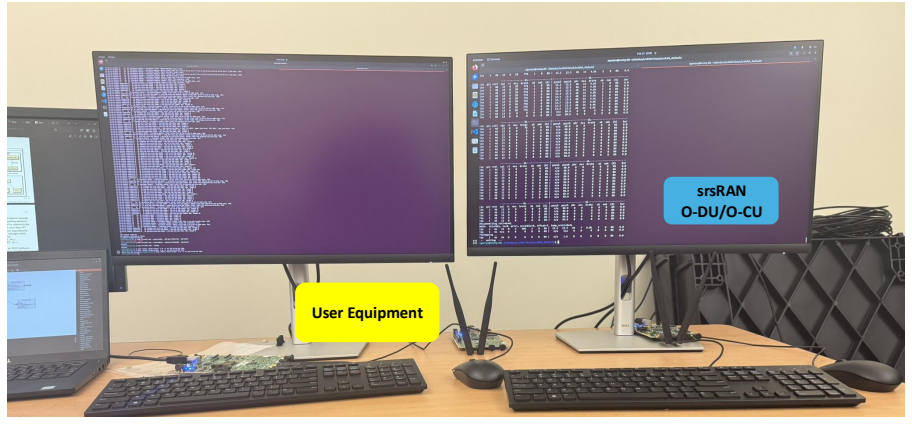

**Figure 3: Testbed setup used for AI5GTest evaluation.**

## 5 Experimental Evaluation

To demonstrate the effectiveness of AI5GTest, we conducted comprehensive experiments using open-source O-RAN implementations. The evaluation focuses on three key aspects: (i) to benchmark the performance of our Gen-LLM against state-of-the-art generative AI models, such as GPT-4o and Gemini; (ii) the accuracy of the Val-LLM and Debug-LLM in identifying deviations and validating multiple test cases; (iii) Timing evaluation of AI5GTest and its constituent components. The following subsections detail the testbed setup, evaluation metrics, and key findings.

### 5.1 O-RAN 5G Testbed Setup

To evaluate the effectiveness of AI5GTest, we deployed a testbed based on the srsRAN Project [34], a widely adopted open-source 5G O-RAN implementation. The testbed supports a range of E2E test cases, covering interactions between the core network, gNB (O-DU, O-CU-CP, O-CU-UP), and user equipment (UE). The srsRAN adopts a monolithic gNB architecture, where the O-CU and O-DU components are integrated into a single entity. A USRP B210 software-defined radio (SDR) serves as the O-RU, enabling over-the-air transmission and reception. Notably, srsRAN provides packet capture files by default during execution, simplifying the Packet Analyzer's task as it can directly process these pre-generated files without additional capture overhead.

The extracted packet files provide a comprehensive and structured representation of signaling interactions in an O-RAN testbed. These files contain fine-grained network information, including protocol-specific messages, configuration parameters, and control plane signaling sequences. While primarily used for validating compliance against O-RAN and 3GPP standards, these structured packet files also have broader applications. They can assist in network troubleshooting, fuzz testing for security vulnerabilities, detailed performance evaluations, and AI-driven anomaly detection tasks. A truncated example of the processed JSON output is provided in **Appendix A.1**. The complete packet files for all executed test cases, along with the full JSON outputs, are available at [1].

### 5.2 Evaluation Metrics

As AI5GTest focuses on automated test procedural flow generation (achieved via Gen-LLM), and the test case validation (achieved through Val-LLM and Debug-LLM), we validate their performance through Gemma-Score and Validation Accuracy, respectively.

*5.2.1 Gemma-Score.* To rigorously evaluate the semantic alignment between the generated procedural flows and their ground truth counterparts from O-RAN and 3GPP specifications, we introduce the *Gemma-Score*, an embedding-based metric inspired by BERTScore [45]. Unlike traditional lexical similarity measures, the Gemma-Score leverages high-dimensional contextual embeddings to assess structural and semantic fidelity. Given a generated procedural flow $G$ and its corresponding ground truth flow $T$, we encode both sequences using the Gemma-2B [36] model, denoted as:

$$\mathcal{E}_{\text{Gemma}} : \mathbb{T} \rightarrow \mathbb{R}^{2048} \tag{1}$$

where $\mathbb{T}$ represents the space of text sequences, and each sequence is mapped to a 2048-dimensional latent space. The embeddings for the generated and ground truth flows are obtained as:

$$\mathbf{E}_G = \mathcal{E}_{\text{Gemma}}(G), \quad \mathbf{E}_T = \mathcal{E}_{\text{Gemma}}(T) \tag{2}$$

To quantify the discrepancy between $G$ and $T$, we compute the Euclidean distance:

$$d(G,T) = \|\mathbf{E}_G - \mathbf{E}_T\|_2 \tag{3}$$

where a lower distance $d(G,T)$ indicates a higher degree of semantic similarity between the generated and ground truth procedural flows. We choose the Gemmna-2B instead of BERT and build upon the BERTScore due to its higher maximum sequence length of 8192 tokens and an output dimension of 2048. Here, a token means a subword or word piece that represents a discrete unit of text; a higher value would indicate a model's ability to address a larger text instance in a single forward pass. This allows Gemma-Score to handle longer sequences of text compared to BERT, which is limited to 512 tokens per sequence and an output dimension of 768 [9]. The output dimension is also an important factor of consideration, as it signifies the degree to which we can embed or represent an input text snippet, with a higher value symbolising an enhanced granularity of comparison.

*5.2.2 Validation Accuracy.* The validation phase treats test case verification as a binary classification task, where a test case is classified as *pass* only if all signaling messages are present and strictly adhere to the chronological order specified in O-RAN/3GPP standards. All other cases—including missing steps and incorrect chronologies—are categorized as *fail*, with the *Partial Pass* category being considered a special case for failure, and prediction is categorized as *fail*. Let $\mathcal{D} = \{(G_i, Y_i)\}_{i=1}^{N}$ represent the performed experiments for assessing the Val-LLM and Debug-LLM, where $G_i$ denotes the $i$-th test case and $Y_i \in \{0,1\}$ indicates its ground truth label (0: fail, 1: pass). The Val-LLM produces predicted labels $\hat{Y}_i \in \{0,1\}$, yielding the following confusion matrix components: *True Positive (TP)* occurs when the ground truth test case is *Pass* ($Y_i = 1$) and the Val-LLM correctly predicts it as *Pass* ($\hat{Y}_i = 1$); *False Positive (FP)* arises when the ground truth test case is *Fail* ($Y_i = 0$), but the Val-LLM incorrectly classifies it as *Pass* ($\hat{Y}_i = 1$); *True Negative (TN)* occurs when the ground truth test case is *Fail* ($Y_i = 0$), and the Val-LLM classify it as *Fail*, and Debug-LLM correctly classify it as *Fail* or *Partial Pass* ($\hat{Y}_i = 0$); Finally, *False Negative (FN)* happens when the ground truth test case is *Pass* ($Y_i = 1$), but the Val-LLM incorrectly classifies it as *Fail* and Debug-LLM incorrectly classify it as *Fail* or *Partial Pass*($\hat{Y}_i = 0$). The Validation Accuracy ($\mathcal{A}_{\text{val}}$) quantifies overall validation correctness by leveraging the TP, FP, TN, and FN with the following formula:

$$\mathcal{A}_{\text{val}} = \frac{TP + TN}{TP + TN + FP + FN} \tag{4}$$

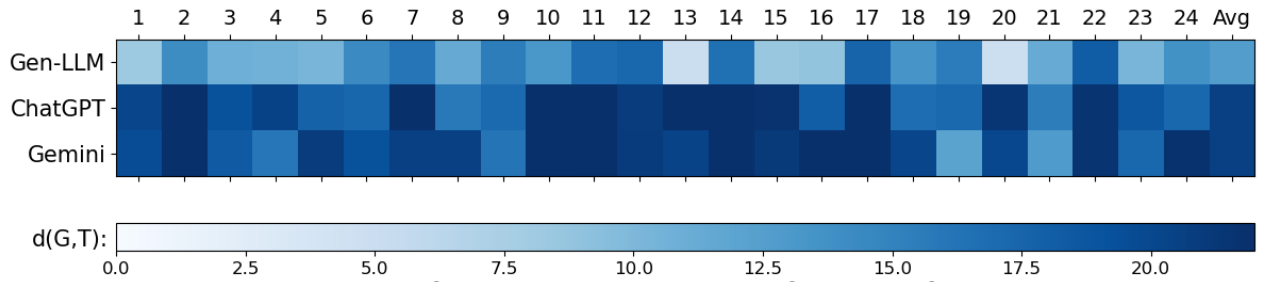

**Figure 4: Heatmap of Gemma-Scores for Different Test Cases. 1-24 represent individual test cases evaluated against ground truth procedural flows.**

## 5.3 Gen-LLM Evaluation

We compare our Gen-LLM to two closed-source LLM solutions, *ChatGPT-4o* [5] and *Gemini-1.5* [35], for test case procedural flow generation. Both these models have been widely regarded as state-of-the-art models in natural language processing, demonstrating strong performance across a variety of benchmarks and real-world applications. These models have demonstrated high performance and have seen increasing adoption in telecommunications use cases [12, 24, 46, 44]. We leverage a total of 24 test cases, as shown in Table 2, to compare our Gen-LLM against ChatGPT and Gemini.

When we consider both ChatGPT and Gemini, we observe that the procedures are generated by significantly relying on the test case name and the components listed in the prompt. While these models often infer a high-level sequence that aligns with the general idea implied by the test case, like initial access, handover, or resource release, they usually hallucinate the technical details present in the specifications. They do generate plausible-sounding message names, such as `RegistrationRequest`, but fail to adhere to the exact sequence mandated by specifications or the correct names. We believe these discrepancies arise because these LLMs lack domain-specific information, and their parametric knowledge is derived from broad corpora rather than targeted exposure to the O-RAN and 3GPP specifications.

Our comparison, as shown in the Figure 4, presents a heatmap visualization of the Gemma-Scores across the 24 test cases. The color intensity represents the $d(G,T)$, where darker shades correspond to higher distances, indicating lower alignment. The heatmap highlights that GPT-4o and Gemini frequently produce high-distance outputs. In contrast, Gen-LLM maintains lower distances across all test cases, reinforcing its reliability in procedural flow generation. Furthermore, our Gen-LLM has the lowest average Gemma-Score of **12.581**; In contrast, GPT-4o achieves an average score of **20.559**, which is 63.4% greater than that of Gen-LLM, while Gemini achieves an average score of **20.579**, which is 63.6% greater compared to that of Gen-LLM.

However, for Gen-LLM we do not observe a perfect Gemma Score $d(G,T) = 0$ every time, and we believe that is due to the Retrieval process as mentioned in Section 4.2 and Appendix A.3. As it is possible to retrieve chunks with a marginally positive reranker score, which can occasionally contain information from related but different test cases, an example being *Initial UE Access* and the *Initial UE Access – UE Context Creation, Service Request* where they have multiple overlapping steps, leading to partially correct procedures. Another scenario that we also observe is that a chunk that contains multiple intermediate steps might get a higher score, due to more relevant content than a chunk with fewer steps (which should precede chronologically), leading to an incorrect ordering in the procedural generation. We believe both of these drawbacks are naturally accounted for by the human-in-the-loop (HITL) nature of the Gen-LLM, ensuring robust and correct procedural generation.

It was also interesting to note that for only one test case that pertains to *gNB-DU Initiated UE Context Release*, we found that a closed-source alternative, Gemini, outperformed the proposed Gen-LLM by obtaining a lower gemma score. Hence, for a direct comparison, Gen-LLM outperforms Gemini for 95.83% of the test cases and outperforms ChatGPT for 100% of the test cases. We further analyzed the performance of the inherent RAG mechanism, with every prompt retrieving relevant documents, and we were able to observe the correct procedural flow for all the test cases in the retrieved chunks. To further showcase the effectiveness of the human-in-the-loop (HITL) system, we consider cases with different ranges of Gemma Score. We observe that 8.33% of samples fall in the 0-5 category, 12.5% in the 5-10 category, 45.83% in the 10-15 category, and the remaining 8 samples are in the 15-20 category.

For the *Initial UE Access* test case, with a Gemma-Score of 4.826, we confirm that the procedural flow exists in the top-ranked document (`38401-fa0.docx`). Similarly, for *Inter gNB-DU Mobility for 5G NSA and SA*, which has a Gemma-Score of 8.240, the required procedural steps are found in Document 1 (`38401-fa0.docx`). Similarly, for *F1 Setup for NR* with a Gemma-Score of 14.137, the relevant procedural flow is also present in Document 1 (`38473-gf0.docx`). We do observe an anomaly for *UE-Initiated Detach Procedure for E-UTRAN* (Gemma-Score of 10.598), where we identify the correct procedural flow in Document rank 3 (`23401-bb0.docx`). We believe this is due to a simple prompting process that we have leveraged, as mentioned in section 4.2. We discuss this limitation in Section 6; however, our evaluation of Gen-LLM demonstrates that by offering users five specifications and implementing the proposed Top-K retrieval methodology, users can effectively leverage Gen-LLM to obtain the correct procedure, an outcome that is not achievable with the currently available closed-source models.

## 5.4 Val-LLM and Debug-LLM Evaluation

To assess the effectiveness of AI5GTest's test case validation, we executed *15* test case instances spanning interoperability and end-to-end (E2E) testing categories as showcased in the table 1. These test cases were implemented through our testbed as shown in section 5.1 under conditions designed to produce a "Pass" outcome. However, through manual scrutiny of execution logs and signaling chronologies, we identified discrepancies, such as misaligned message sequences and non-compliance with the 3GPP and O-RAN standards. Each test case instance was manually labeled as *Pass*, *Fail*, or *Partial Pass* based on observed behavior (see Section 4). Subsequently, all labeled instances were validated using the Val-LLM module, and failures were further analyzed by the Debug-LLM module to isolate root causes. The results of this validation process are detailed in Table 1.

Our proposed validation framework exhibited remarkable precision in verifying test cases and across the 15 scenarios that we validated, we obtain seven $TP$, eight $TN$ and zero $FP$, $FN$, resulting in a $\mathcal{A}_{\text{val}}$ accuracy of 100%. Additionally, the primary inconsistencies leading to a *Partial Pass label* were observed in test cases related to *Initial UE Access*, where a recurring chronological error caused the signaling message between gNB-CU and AMF—intended to execute at the end—to appear prematurely. For a better understanding of the Val-LLM and Debug-LLM's performance for $TN$ instance, we provide a comprehensive visualization for the *UE Initial Access E1 over F1* test case, where we were able to identify the most discrepancies with the proposed method accurately detecting them all. The Figure 5 showcases the VAL-LLM and Debug-LLM execution, and a detailed tracing and explanation of the same is available in the Appendix A.2.

**Table 1: Validation (Val-LLM and Debug-LLM) Results.**

| TC Num. | Test Case Title | Ground Truth | Val-LLM | Debug-LLM | Inference |
|---|---|---|---|---|---|
| **TC-01** | Initial UE access – UE Context Creation, Service Request. | Partial Pass | Fail | Partial Pass | Incorrect chronology in signaling sequence (The signalling message where gNB-CU should send an `INITIAL CONTEXT SETUP RESPONSE` to the AMF is executed prematurely) |
| **TC-02** | Initial access – UE Context Creation for Initial Registration. | Partial Pass | Fail | Partial Pass | Incorrect chronology in signaling sequence (The signalling message where gNB-CU should send a `UL NAS TRANSPORT` (Registration Complete) to the AMF is executed prematurely) |
| **TC-03** | Registration Update without Follow-on Request. | Pass | Pass | - | - |
| **TC-04** | gNB-CU Initiated UE Context Modification. | Pass | Pass | - | - |
| **TC-05** | gNB-DU Initiated UE Context Release. | Pass | Pass | - | - |
| **TC-06** | FI Setup for NR. | Pass | Pass | - | - |
| **TC-07** | UE Initial Access over F1 | Partial Pass | Fail | Partial Pass | Incorrect chronology in signaling sequence (The signalling message where gNB-CU should send an `INITIAL CONTEXT SETUP RESPONSE` to the AMF is executed prematurely) |
| **TC-08** | Bearer Context Setup over F1-U. | Pass | Pass | - | - |
| **TC-09** | RRC Connected to RRC Inactive. | Pass | Pass | - | - |
| **TC-10** | PDU Session Establishment. | Fail | Fail | Fail | Incorrect message name (`PDU SESSION RESOURCE SETUP REQUEST` instead of `PDU SESSION RESOURCE REQUEST`) |
| **TC-11** | UE Initial Access over E1 and F1. | Partial Pass | Fail | Partial Pass | Incorrect chronology in signaling sequence (Multiple discrepancies in the expected chronology) |
| **TC-12** | gNB-CU-UP Initiated Bearer Context Release over F1-U. | Pass | Pass | - | - |
| **TC-07*** | UE Initial Access over F1: Simulated Failure (5GC Crash) | Fail | Fail | Fail | Both algorithms converge early due to the absence of the `INITIAL CONTEXT SETUP REQUEST` message from the AMF to the gNB-CU, leading to an incomplete signaling procedure |
| **TC-07*** | UE Initial Access over F1: Simulated Failure (IMSI Mismatch) | Fail | Fail | Fail | Both algorithms converge early due to the absence of the `INITIAL CONTEXT SETUP REQUEST` message from the AMF to the gNB-CU, leading to an incomplete signaling procedure |
| **TC-07*** | UE Initial Access over F1: Simulated Failure (USIM Algo mismatch) | Fail | Fail | Fail | Both algorithms converge early due to the absence of the `INITIAL CONTEXT SETUP REQUEST` message from the AMF to the gNB-CU, leading to an incomplete signaling procedure |

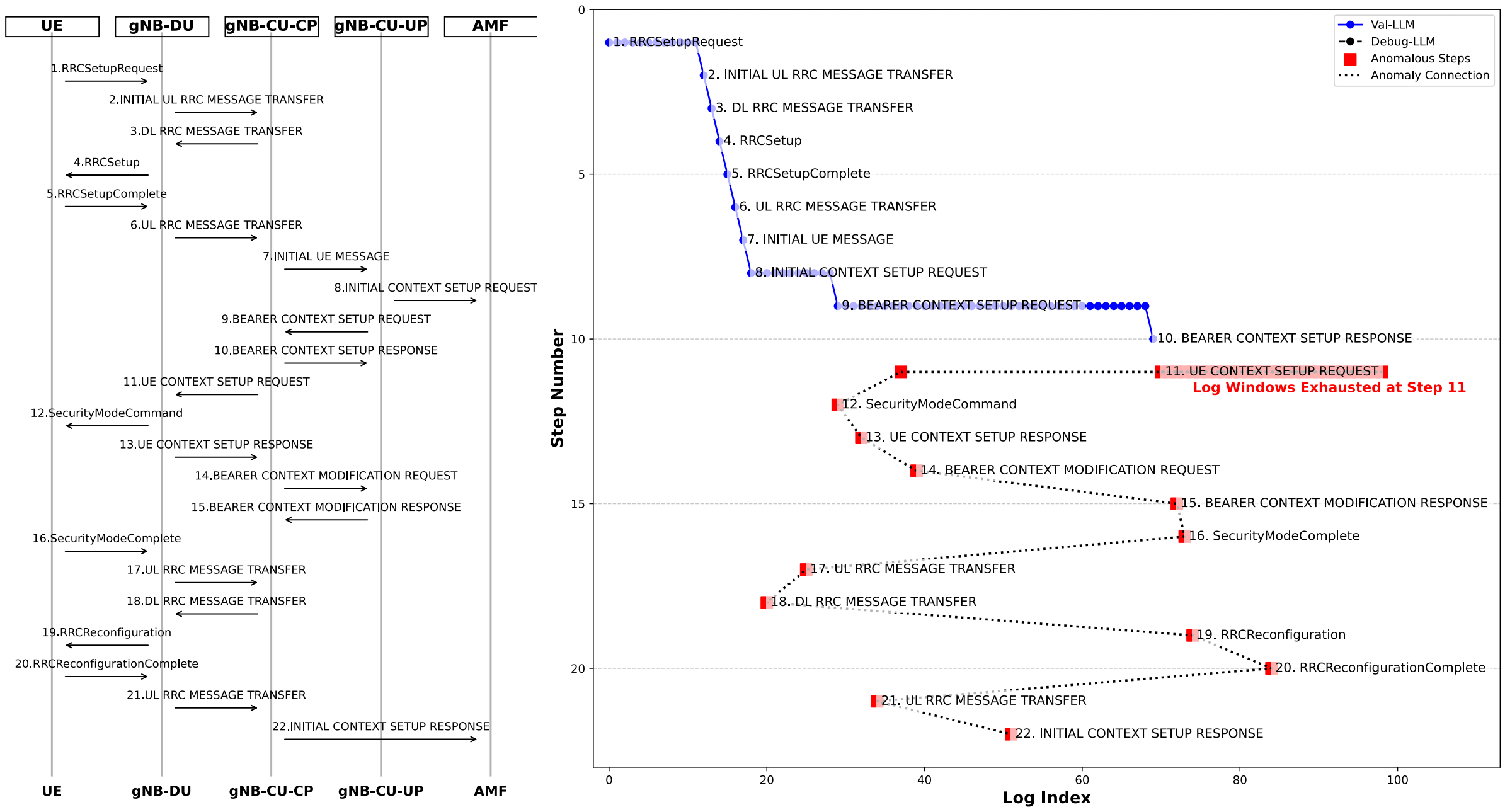

**Figure 5: Visualization of the *UE Initial Access over E1 and F1* test case validation with the Ground label Fail (Partial Pass). The Val-LLM converges at Step 11, and the remaining log processing is obtained through the Debug-LLM. The correct procedural flow, as mentioned in 3GPP TS-Spec 38.401, is mentioned in the left-hand figure.**

## 5.5 Timing Evaluation

Based on the available computational resources, we perform the RAG operations for the Gen-LLM locally with the hardware configuration consisting of an Intel(R) Core(TM) i9-14900KF CPU with 62 GB of RAM, paired with an NVIDIA GeForce RTX 4090 GPU with 24 GB of GDDR6X memory. The Mistral model in Gen-LLM and the $LLaMA$ calls for the Val-LLM and Debug-LLM are performed using `langchain-nvidia-ai-endpoints` [20].

The Gen-LLM inference takes an average of 33.443 seconds with a standard deviation of 0.791 seconds. If we strictly consider the *Pass instances*, it takes an average of 3.06 minutes with a standard deviation of 35.312 seconds to execute the Val-LLM algorithm. As Debug-LLM is only triggered for *Fail* test cases, for the 8 $TN$ instances, we obtain an average execution time of 41.873 minutes with a standard deviation of 8.053 minutes for a Debug-LLM run. Hence, the cumulative average across all validation experiments is 47.994 minutes, and if we consider the complete pipeline, it only takes **48.551 minutes** for AI5GTest to address a test case.

## 5.6 AI5GTest Evaluation

Traditional O-RAN testing remains heavily manual and resource-intensive, requiring engineers to extract expected procedural flows from multiple standard documents, execute test cases, and validate results by manually cross-referencing signaling logs. According to industry reports, manual validation of O-RAN components typically spans multiple weeks to months. In a 16-week O-RAN validation pilot, Aspire Technology tested 45 test cases, averaging approximately 15 hours per test case execution [37]. In contrast, AI5GTest significantly reduces execution time, completing a full test cycle—including procedural flow generation, validation, and debugging—in under an hour per test case (see Section 5.5). This multi-fold reduction in testing duration enables large-scale automation, making O-RAN validation both scalable and repeatable. Beyond efficiency, AI5GTest enhances test reliability and accuracy. Manual validation is inherently prone to human errors, inconsistencies, and vendor-specific interpretations of O-RAN test cases, leading to variability in validation outcomes. AI5GTest eliminates this subjectivity by leveraging domain-specific LLMs, ensuring that validation is conducted with strict adherence to O-RAN and 3GPP standards. Additionally, AI5GTest efficiently handled 24 test cases, including both procedural flow generation and validation, while also identifying failure scenarios such as 5GC crashes, USIM algorithm mismatches, and UE IMSI inconsistencies. These results demonstrate AI5GTest's capability to streamline conformance and interoperability validation at scale, minimizing human intervention while maintaining alignment with industry standards.

## 5.7 Discussion

While our evaluation of AI5GTest and its key components (Gen-LLM, Val-LLM, and Debug-LLM) has been primarily conducted using a single open-source cellular stack (srsRAN), the framework is fundamentally designed to be agnostic to the underlying 5G platform. This platform-independence is achieved through AI5GTest's reliance on standardized 3GPP protocol layers, combined with interface-level packet capture and protocol dissection. As long as a 5G system can export PCAP traces, AI5GTest can operate without requiring any modifications to the system under test. Furthermore, both the procedural flow generation and signaling validation in AI5GTest are driven by 3GPP specifications, not implementation-specific behavior. This enables the framework to generalize across different platforms, including other open-source stacks such as OpenAirInterface, as well as commercial, multi-vendor 5G systems that adhere to standardized interface behavior. This extensibility underscores AI5GTest's potential to serve as a drop-in, automated testing solution for heterogeneous, multi-vendor 5G environments.

# 6 Conclusion and Future Work

This paper introduces AI5GTest – a novel AI-powered, specification-aware testing framework that automates procedural flow generation, signaling validation, and root cause analysis. By leveraging the Gen-LLM, Val-LLM, and Debug-LLM, AI5GTest significantly reduces manual effort and enhances test consistency and scalable validation of O-RAN implementations. Our evaluation demonstrates that Gen-LLM outperforms state-of-the-art GenAI models in accurately generating procedural flows that align with O-RAN and 3GPP specifications. Additionally, AI5GTest provides automated validation and debugging capabilities, offering a faster and more reliable alternative to manual testing processes. An exciting future direction is automated test case generation using LLMs. Currently, AI5GTest relies on predefined test cases from O-RAN WGs, but expanding its scope to generate new, uncovered edge-case test scenarios could improve the detection of unforeseen issues in O-RAN deployments. We also believe that working towards sophisticated prompting techniques that enhance the TC formatter with context would help guarantee that the relevant document is always retrieved at Rank 1. As O-RAN adoption continues to grow, AI5GTest paves the way for scalable, intelligent, and automated testing solutions, ensuring more efficient, consistent, and comprehensive validation of 5G/NextG cellular networks.

# 7 Acknowledgment

Authors acknowledge the funding support from the Public Wireless Supply Chain Innovation Fund (PWSCIF) under Federal Award ID Numbers 26-60-IF010 and 51-60-IF007. We also thank the anonymous shepherd and reviewers for their constructive feedback, which greatly contributed to improving the quality of this paper.

# A Appendix

## A.1 Sample Log Index and *LLaMA*

We test a positive instance for the *Initial UE Access over F1* test case while checking Step 1, *The gNB-DU sends a* `UE CONTEXT RELEASE REQUEST` *to the gNB-CU, indicating the need to release the UE context.*

```
{
 'user_dlt': [],
 'f1ap': [
  'F1AP-PDU: initiatingMessage (0)',
  'initiatingMessage',
  'procedureCode: id-UEContextReleaseRequest (10)',
  'criticality: ignore (1)',
  'value',
  'UEContextReleaseRequest',
  'protocolIEs: 3 items',
  'Item 0: id-gNB-CU-UE-F1AP-ID',
  'ProtocolIE-Field',
  'id: id-gNB-CU-UE-F1AP-ID (40)',
  'GNB-CU-UE-F1AP-ID: 0',
  'GNB-DU-UE-F1AP-ID: 0',
  'Cause: radioNetwork (0)',
  'radioNetwork: rl-failure-others (12)',
  'criticality: reject (0)',
  'criticality: reject (0)',
  'criticality: ignore (1)',
  'value',
  'value',
  'value',
  'Item 1: id-gNB-DU-UE-F1AP-ID',
  'Item 2: id-Cause',
  'ProtocolIE-Field',
  'ProtocolIE-Field',
  'id: id-gNB-DU-UE-F1AP-ID (41)',
  'id: id-Cause (0)'
 ]
}
```

**Label:**

**Yes**

**Confidence Score:**

**100%**

**Explanation:**

The log file contains the exact message name `"UEContextReleaseRequest"` which matches the message name in the test case step. Additionally, the log file indicates that the message is sent from the gNB-DU to the gNB-CU, as it mentions `"f1ap"` which corresponds to communication between the gNB-DU and gNB-CU. The presence of `"GNB-CU-UE-F1AP-ID"` and `"GNB-DU-UE-F1AP-ID"` in the log file further validates that the communication occurs between the correct entities. Therefore, the step has been executed as described in the test case.

## A.2 *TN*: UE Initial Access over E1 and F1

The packet file obtained from the PCAP Analyzer contains a total of 99 log indices, and the test case *UE Initial Access over E1 and F1* consists of 22 distinct steps, with the procedural flow shown in figure 5. The validation process starts with the Val-LLM and Step 1, where the algorithm is supposed to start with identifying the execution of the first step *The UE sends an* `RRCSetupRequest` *to the gNB-DU* Beginning at log index 0, the model computes confidence scores for each index's alignment with the step description. After 11 unsuccessful attempts (indices 0–10, all scoring Label=No with Confidence=0), the algorithm detects a match at index 11. Here, the *LLaMA* forward pass identifies the RRCSetupRequest message (Label=Yes, Confidence Score=100) exchanged between the UE and gNB-DU. The execution triggers a state transition to Step 2. Subsequent steps follow a similar pattern: the Val-LLM advances through log indices until the expected message sequence is validated. For instance, Step 2: *The gNB-DU forwards an* `INITIAL UL RRC MESSAGE TRANSFER` *to the gNB-CU-CP* is immediately confirmed at log index 12 (Confidence=100). Notably, some steps require scanning multiple indices (e.g., Step 8 spans 18 indices before validation at log index 28), reflecting dynamic timing variations in signaling exchanges. The algorithm terminates if a step exhausts all candidate indices without a match, as seen in Step 11, which failed to validate across 29 indices (69–98). Hence, Val-LLM converges with the label *Fail*, and the first phase of the validation process is completed.

**Table 2: Test Case Details.**

| TC Number | Test Category | Test Case Title | DUT | Referenced Documents |
|---|---|---|---|---|
| TC-01 | Interoperability, E2E | Initial UE access – UE Context Creation, Service Request. | O-CU, O-DU | O-RAN NR C-Plane Profile Spec - Clause 6.1.1 |
| TC-02 | Interoperability, E2E | Initial access – UE Context Creation for Initial Registration. | O-CU, O-DU | O-RAN NR C-Plane Profile Spec - Clause 6.1.2 |
| TC-03 | Interoperability, E2E | Registration Update without Follow-on Request. | O-CU, O-DU | O-RAN NR C-Profile Spec v13 - Clause 6.1.3 |
| TC-04 | Interoperability, E2E | gNB-CU Initiated UE Context Modification. | O-CU, O-DU | O-RAN NR C-Profile Spec v13 – Clause 6.3.1.1 |
| TC-05 | Interoperability, E2E | gNB-DU Initiated UE Context Release. | O-CU, O-DU | O-RAN NR C-Profile Spec v13 – Clause 6.2.2 |
| TC-06 | Interoperability | FI Setup for NR SA. | O-CU, O-DU | O-RAN NR C-Profile Spec v13 – Clause 4.2.3.1 |
| TC-07 | Interoperability, E2E | UE Initial Access over F1. | O-CU, O-DU | 3GPP TS-Spec 38.401 v18 – Clause 8.1 |
| TC-08 | Interoperability, E2E | Bearer Context Setup over F1-U. | O-CU-CP, O-CU-UP, O-DU | 3GPP 38.401 v18 – Clause 8.9.2 |
| TC-09 | Interoperability, E2E | RRC Connected to RRC Inactive. | O-CU, O-DU | O-RAN NR C-Profile Spec v13 - Clause 6.9.1 |
| TC-10 | Interoperability, E2E | PDU Session Establishment. | O-CU, O-DU | O-RAN NR C-Profile Spec v13 - Clause 6.3.3.1 |
| TC-11 | Interoperability, E2E | UE Initial Access over E1 and F1. | O-CU-CP, O-CU-UP, O-DU | 3GPP TS-Spec 38.401 v18 – Clause 8.9.1 |
| TC-12 | Interoperability, E2E | gNB-CU-UP Initiated Bearer Context Release over F1-U. | O-CU-CP, O-CU-UP, O-DU | 3GPP TS-Spec 38.401 v18 – Clause 8.9.3.1 |
| TC-13 | E2E - Functional | LTE/5G NSA Attach of a Single UE | EPC, gNB | 3GPP TS 23.401 - Clause 5.3.2.1 |
| TC-14 | E2E - Functional | LTE/5G NSA Detach of a Single UE | EPC, gNB | 3GPP TS 23.401 - Clause 5.3.8.2.1 |
| TC-15 | E2E - Functional | LTE/5G NSA Attach of Multiple UEs | EPC, gNB | 3GPP TS 23.401 - Clause 5.3.2.1 |
| TC-16 | E2E - Functional | LTE/5G NSA Detach of Multiple UEs | EPC, gNB | 3GPP TS 23.401 - Clause 5.3.8.2.1 |
| TC-17 | E2E - Functional | General Registration of UE in 5G SA | 5GC, gNB | 3GPP TS 23.502 - Clause 4.2.2.2.2 |
| TC-18 | Interoperability, E2E | Intra O-DU Mobility for 5G NSA and SA | O-CU, O-DU | 3GPP TS 38.401 - Clause 8.2.1 |
| TC-19 | Interoperability, E2E | Inter O-DU Mobility for 5G NSA and SA | O-CU, O-DU | 3GPP TS 38.401 - Clause 8.2.1 |
| TC-20 | Interoperability, E2E | Inter O-CU Mobility for 5G NSA and SA | O-CU-CP, O-CU-UP, O-DU | 3GPP TS 38.401 - Clause 8.9.4 (SA); 3GPP TS 37.340 - Clause 10.5.1 (NSA) |
| TC-21 | E2E - Functional | Registration to a Single emBB Network Slice in 5G SA | 5GC, gNB | 3GPP TS 23.502 - Clause 4.2.2.2.2; TIFG.E2E-Test Spec-Table 5.7 |
| TC-22 | E2E - Functional | De-registration from a Single emBB Network Slice in 5G SA | 5GC, gNB | 3GPP TS 23.502 - Clause 4.2.2.3.2; TIFG.E2E-Test Spec-Table 5.8 |
| TC-23 | E2E - Functional | Registration to Multiple Network Slices in 5G SA | 5GC, gNB | 3GPP TS 23.502 - Clause 4.2.2.2.2; TIFG.E2E-Test Spec-Table 5.10 |
| TC-24 | E2E - Functional | De-registration from Multiple Network Slices | 5GC, gNB | 3GPP TS 23.502 - Clause 4.2.2.3.2; TIFG.E2E-Test Spec-Table 5.11 |

This triggers the Debug-LLM to further compute the heirarchical log processing. Here, as for a particular step, we execute *LLaMA* forward passes iteratively for all log indices regardless of the execution trends of the previous steps; we can identify if a step was executed in the entirety of the 99 log indices. For example, during reprocessing, Step 19 (*The UE responds with an* `RRCReconfigurationComplete` *message to the gNB-DU*) is detected at log index 74 (Confidence=100), despite preceding steps (e.g., Step 18 at log index 20 and Step 17 at log index 25) occurring earlier in the trace. Notably, all messages are tracked using a `used-indices` set, like for Steps 6, 17, and 21, which technically involve the same signaling message (e.g., `INITIAL UL RRC MESSAGE TRANSFER`) between the UE and gNB-DU. This ensures their log indices (e.g., 25 for Step 17, 34 for Step 21) are not redundantly processed, avoiding overlapping interpretations of the same procedural instance. These deviations highlight protocol timing mismatches and errors in implementation. By aggregating such anomalies (e.g., Step 20 at log index 84, Step 21 at 34, and Step 22 at 51), the Debug-LLM reconstructs the procedural flow, identifying disjointed signaling sequences. The results align with the hybrid execution trends plotted in Figure 5, where steps are mapped to log indices non-linearly.

## A.3 Reranked Chunks

**Score > 4 (4.100)**

**Source**: *38401-fa0.md*

...(1) The UE sends an `RRCSetupRequest` message to the gNB-DU.
(2) The gNB-DU includes the RRC message and, ...low layer configuration for the UE in the `INITIAL UL RRC MESSAGE TRANSFER` ...
(8) The AMF sends the `INITIAL CONTEXT SETUP REQUEST` message to the gNB-CU. ...

**1 ≥Score > 0 (0.851)**

**Source**: *38473-gf0.md*

The establishment of the UE-associated logical F1 connection shall be initiated as part of the procedure ...If the SUL Access Indication IE is included in the `INITIAL UL RRC MESSAGE TRANSFER`, the gNB-CU shall consider that the UE has performed ...

**0 ≥Score > -1 (-0.688)**

**Source**: *38473-h60.md*

```
Direction: gNB-CU → gNB-DU
---------------------------------------------
IE/Group Name              Presence  Range
Message Type               M         9.3.1.1 .....
gNB-CU UE F1AP ID          M         9.3.1.4 .....
gNB-DU UE F1AP ID          M         9.3.1.5 .....
RAN UE PDC Measurement ID M          INTEGER .....
---------------------------------------------
```

**Score ≤-4 (-6.994)**

**Source**: *33846-h00.md*

| | | | | |
|---|---|---|---|---|
| | USIM:<br>USIM | | | to UE and<br>UDM only. |

Here we have illustrated snippets of the retrieved chunks across 4 different score ranges, and we do observe that positive reranker scores especially very high values (usually > 4) are usually extremely important for generating the correct procedure, but when we look at chunks with a marginally positive score (0-1) there are cases like the one mentioned below where we get tangential information and sometimes we can also observe information regarding neighboring test cases which have some overlapping steps. Without a reranker, this conflicting information can easily confuse the `Mistral-7B-Instruct`, leading to partially incorrect generations with some incorrect steps or an incorrect ordering of the steps.